\documentclass[aps,pra,showpacs,preprint]{revtex4}

\usepackage{amsfonts}
\usepackage{amsmath}
\usepackage{amssymb}
\usepackage{graphicx}
\usepackage[T2A]{fontenc}
\usepackage[cp1251]{inputenc}
\usepackage{srcltx}%

\def\figw{7cm}

\begin{document}


\title{Averaged optical characteristics of an ensemble of metal nanoparticles}


\author{P.M. Tomchuk}
\author{V.N. Starkov}
\email[Corresponding author: ]{vjachnikstar@gmail.com}

\affiliation{Institute of Physics, National Academy of Sciences of Ukraine, Department of
Theoretical Physics,\\46 Nauka Ave., Kyiv, 03680, Ukraine}


\date{\today}

\begin{abstract}
A theory for the averaged optical characteristics of an ensemble of metal
nanoparticles with different shapes has been developed. The theory is
applicable both for the nanoparticle size at which the optical conductivity of
the particle is a scalar and for the nanoparticle size at which the optical
conductivity should be considered as a tensor. The averaged characteristics
were obtained taking into account the influence of nanoparticle shape on the
depolarization coefficient and the components of the optical conductivity
tensor. The dependences of magnetic absorption by a spheroidal metal
nanoparticle on the ratio between its curvature radii and the angle between
the spheroid symmetry axis and the magnetic field vector were derived and
theoretically considered. An original variant of the distribution function for
nanoparticle shapes, which is based on the combined application of the
Gaussian and \textquotedblleft hat\textquotedblright\ functions, was proposed
and analyzed.
\end{abstract}


\maketitle


\section{Introduction}

A typical attribute of the optical spectra of ensembles of metal nanoparticle
is the availability of plasma resonances. The number of those resonances, as
well as their frequency positions and damping decrements, depend on the metal
nanoparticle shape (see, e.g., Refs.~\cite{1,2}). Since it is hardly possible
to create an ensemble of absolutely identical nanoparticles, experimenters,
when studying the processes of light absorption and scattering by such
ensembles, usually deal with averaged (apparent) optical characteristics. The
procedure of averaging the optical characteristics for ensembles of spheroidal
metal nanoparticles was described, e.g., in Ref\.{s}.~\cite{3,4}.

While examining the influence of the shape of metal nanoparticles on their
optical properties, the authors of previous papers supposed that dissipative
processes in nanoparticles are characterized by a scalar high-frequency
conductivity. However, we have shown \cite{2,5,6} that if the size of
non-spherical nanoparticles does not exceed the electron free path length, the
optical conductivity becomes a tensor quantity. The diagonal elements of this
tensor together with the depolarization coefficients govern the half-widths of
plasma resonances \cite{2,7}. In this case, the averaging over the
nanoparticle shape is no more reduced to the averaging over the depolarization
coefficients \cite{3}. Therefore, in this paper, we consistently summarized
the theoretical basis that was previously developed for the optical properties
of an ensemble of elliptic metal nanoparticles, including the components of
the optical conductivity tensor and the depolarization tensor \cite{2,5,6,7}.
As a result, the depolarization coefficients and the components of the optical
conductivity tensor averaged over the nanoparticle forms were calculated.
While averaging over the nanoparticle shapes, the influence of this parameter
on the conductivity was taken into account for the first time.

\section{Formulation of the problem}

The study of optical properties of nanoparticles has a long history (see,
e.g., Refs. \cite{1,3,8,9}). In particular, the expression for the absorption
cross-section of a plane electromagnetic wave with the frequency $\omega$ by a
spherical nanoparticle which size is smaller than the wavelength has been
known for a long time~\cite{10}:
\begin{equation}
K(\omega)=\frac{12\pi\,\omega\,a^{3}\epsilon^{\prime\prime}}{c}\left\{
{\frac{1}{|\epsilon|^{2}}+\frac{\omega^{2}a^{2}}{90\,c^{2}}}\right\}
.\label{eq1}%
\end{equation}
Here, $c$ is the light {velocity}; $\epsilon$ the dielectric permittivity,
which has the form
\begin{equation}
\epsilon=\epsilon^{\prime}+i\epsilon^{\prime\prime}=1-\frac{\omega_{p}^{2}%
}{\omega^{2}+\nu^{2}}+i\,\frac{\nu}{\omega}\frac{\omega_{p}^{2}}{\omega
^{2}+\nu^{2}}\label{eq2}%
\end{equation}
in the Drude model \cite{1}; $\nu$ is the bulk collision frequency;
\begin{equation}
\omega_{p}=\left(  \frac{4\pi n_{0}e^{2}}{m}\right)  ^{1/2}\label{eq3}%
\end{equation}
is the plasma frequency; and $e$, $m$, and $n_{0}$ are the electron charge,
mass, and concentration, respectively. The first term in the braces in
Eq.~(\ref{eq1}) is associated with the electric wave component, and the
corresponding absorption is called electric absorption, whereas the second
term is associated with the magnetic component and the corresponding
absorption is called magnetic \cite{10}.

The most general theory describing the optical properties of small particles,
which simultaneously is the most cited one, is the Mie theory \cite{11}. It
was developed for spherical particles in the assumption that the vector of
electric current density in the particle, $\mathbf{{j}}(\mathbf{{r}},t)$, is
related to the generating field $\mathbf{{E}}(\mathbf{{r}},t)$ by Ohm's law
\begin{equation}
\mathbf{{j}}(\mathbf{{r}},t)=\sigma\,\mathbf{{E}}(\mathbf{{r}},t), \label{eq4}%
\end{equation}
where $\sigma$ is the scalar conductivity, $\mathbf{r}$ the coordinate vector,
and $t$ the time. Generally speaking, relation (\ref{eq4}) is valid for
particles which sizes are much larger than the free electron path length
\cite{2}. Otherwise, this relation becomes non-local \cite{12}, and, moreover,
the conductivity becomes a tensor quantity in the case of asymmetric particles
\cite{2}.

Let a metal particle be in the field of an external electromagnetic wave
\begin{equation}
\left(  {%
\begin{array}
[c]{l}%
\mathbf{{E}}\\
\mathbf{{H}}%
\end{array}
}\right)  =\left(  {%
\begin{array}
[c]{l}%
\mathbf{{E}}^{(0)}\\
\mathbf{{H}}^{(0)}%
\end{array}
}\right)  \;e^{i\left(  {\mathbf{{k}}\mathbf{{r}}-\omega t}\right)  },
\label{eq5}%
\end{equation}
where $\mathbf{E}$ and $\mathbf{H}$ are the electric and magnetic,
respectively, components of the wave field; and $\mathbf{k}$ is the wave
vector ($k=2\pi/\lambda$, where $\lambda$ is the wavelength). Then, the
problem of finding the current density vector $\mathbf{{j}}(\mathbf{{r}},t)$,
which governs the energy absorption, is split into two stages. At the first
stage, we have to determine internal fields generated by wave (\ref{eq5}) in
the nanoparticle. At the second stage, we have to determine how these internal
fields modify the electron {velocity }distribution function, i.e. to find a
correction induced by the internal fields to the equilibrium Fermi distribution.

Since the internal fields induced by wave (\ref{eq5}) in the nanoparticle
depend on the particle shape, we will assume that the nanoparticle is an
ellipsoid. It is convenient to develop a theory for this form, because the
results obtained can be extended to a wide range of nanoparticle forms (from
discoid- to rod-shaped) by changing the ellipsoid curvature radii.

If the wavelength $\lambda$ is much larger than the nanoparticle size
($\lambda\gg\max\left\{  {R_{i}}\right\}  $, where $R_{i}~(i=1,2,3)$ are the
curvature radii), then the relation between the internal fields and external
ones ($\mathbf{{E}}^{(0)}$ and $\mathbf{{H}}^{(0)}$) is known \cite{10}. In
particular, in the coordinate frame oriented along the principal ellipsoid
axes, the electric (potential) component of the internal field has the form
\cite{10}
\begin{equation}
(E_{in})_{j}=\frac{E_{j}^{(0)}}{1+L_{j}\left[  \epsilon{(\omega)-1}\right]
},\label{eq6}%
\end{equation}
where $L_{j}$ is the $j$-th diagonal component of the depolarization tensor.
Similarly, the eddy component of the electric field induced by the magnetic
field $\mathbf{{H}}^{(0)}$ equals \cite{2}
\begin{equation}
(E_{ed})_{x}=i\frac{\,\omega\,}{c}\left\{  {\frac{z\,H_{y}^{(0)}}{R_{z}%
^{2}+R_{y}^{2}}-\frac{y\,H_{x}^{(0)}}{R_{x}^{2}+R_{y}^{2}}}\right\}
.\label{eq7}%
\end{equation}
The other eddy field components can be obtained from Eq.~(\ref{eq7}) by
cyclically permutating the subscripts. Let us also recall that $H^{(0)}%
=E^{(0)}$.

\section{Electron {velocity }distribution function and energy absorption by
metal nanoparticles}

The velocity distribution function of electrons in metal nanoparticles in the
presence of fields (\ref{eq6}) can be expressed as the sum of the equilibrium
Fermi function $f_{0}(\varepsilon)$, where $\varepsilon=\frac{1}{2}mv^{2}$ is
the electron energy, and the solution $f_{1}(\mathbf{{r}},\mathbf{{v}})$ of
the kinetic equation linearized in the total (potential and eddy) field
\begin{equation}
\mathbf{{F}}=\mathbf{{E}}_{in}+\mathbf{{E}}_{ed}; \label{eq9}%
\end{equation}
namely,
\begin{equation}
(\nu-i\,\omega)\,f_{1}+\mathbf{{v}}\frac{\partial f_{1}}{\partial\mathbf{{r}}%
}+e\,\mathbf{{F}}(\mathbf{{r}})\mathbf{{v}}\frac{\partial f_{0}}%
{\partial\varepsilon}=0. \label{eq8}%
\end{equation}
The function $f_{1}(\mathbf{{r}},\mathbf{{v}})$ must also satisfy the boundary
condition%
\begin{equation}
\left.  {f_{1}(\mathbf{{r}},\mathbf{{v}})}\right\vert _{S}=0~\mathrm{at~}%
v_{n}<0, \label{9}%
\end{equation}
where $v_{n}$ is the {velocity} component directed normally to the surface.
The solution of the boundary problem (\ref{eq8}), (\ref{9}) for the
ellipsoidal particle looks like
\begin{equation}
f_{1}(\mathbf{{r}},\mathbf{{v}})=-{f}_{0}^{\prime}(\varepsilon)\int
\limits_{0}^{t_{0}}{d\tau\,e^{-\tilde{{\nu}}(t_{0}-\tau)}e\,\mathbf{{F}%
}({\mathbf{{r}}}^{\prime}-{\mathbf{{v}}}^{\prime}(t_{0}-\tau))}. \label{eq10}%
\end{equation}
Here, $\tilde{{\nu}}=\nu-i\,\omega$, and
\begin{equation}
t_{0}=\frac{1}{{v}^{\prime}{}^{2}}\left[  {\mathbf{{r}}}^{\prime}{\mathbf{{v}%
}}^{\prime}+\sqrt{(R^{2}-{r}^{\prime}{}^{2}){v}^{\prime}{}^{2}+({\mathbf{{r}}%
}^{\prime}{\mathbf{{v}}}^{\prime})^{2}}\right]  \label{eq11}%
\end{equation}
is a characteristic of Eq.~(\ref{eq8}). The primes in Eqs.~(\ref{eq10}) and
(\ref{eq11}) mark the corresponding quantities in a deformed coordinate frame,
in which the ellipsoidal particle becomes spherical \cite{2}. The coordinate
and {velocity components }in the deformed and undeformed frames are mutually
related:
\begin{equation}
x_{j}=\frac{R_{j}}{R}{x}_{j}^{\prime},\quad v_{j}=\frac{R_{j}}{R}{v}%
_{j}^{\prime},\quad(j=1,2,3), \label{eq12}%
\end{equation}
where $R=(R_{1}R_{2}R_{3})^{1/3}$.

Knowing the electron {velocity }distribution function, the current density
vector can be found:
\begin{equation}
\mathbf{{j}}(\mathbf{{r}},\omega)=\mathrm{Re}\left(  {\frac{m}{2\pi\hbar}%
}\right)  ^{3}\int{\mathbf{{v}}\,f_{1}(\mathbf{{r}},\mathbf{{v}})}%
\,d^{3}(v).\label{eq13}%
\end{equation}
In accordance with Eqs.~(\ref{eq9}) and (\ref{eq10}), this vector has two
components: the electric and magnetic ones induced by $\mathbf{{E}}_{in}$ and
$\mathbf{{E}}_{ed}$, respectively:
\begin{equation}
\mathbf{{j}}(\mathbf{{r}},\omega)=\mathbf{{j}}_{e}(\mathbf{{r}},\omega
)+\mathbf{{j}}_{m}(\mathbf{{r}},\omega),\label{eq14}%
\end{equation}
The explicit forms can be determined for both components by substituting
Eq.~(\ref{eq10}) into Eq.~(\ref{eq13}).

Accordingly, the energy absorbed by the metal nanoparticle,
\begin{equation}
W=W_{e}+W_{m}=\frac{1}{2}\mathrm{Re}\int\limits_{V}{d\,\mathbf{{r}}}\left\{
{\mathbf{{j}}_{e}\,\mathbf{{E}}_{in}^{\ast}+\mathbf{{j}}_{m}\,\mathbf{{E}%
}_{ed}^{\ast}}\right\}  , \label{eq15}%
\end{equation}
is also the sum of the electric and magnetic contributions. By normalizing
Eq.~(\ref{eq15}) to the magnitude of the flux incident on the nanoparticle, we
obtain the absorption coefficients. For example, the coefficient of electric
light absorption by a metal particle of the volume $V$ located in a matrix
with the dielectric constant $\epsilon_{m}$ equals{}%
\begin{equation}
K\left(  \omega\right)  =4\pi\,\mathrm{Re}\int\limits_{V}{d\mathbf{{r}}}%
\frac{{\mathbf{{j}}_{e}\left(  {\mathbf{{r}},\omega}\right)  \mathbf{{E}}%
_{in}^{\ast}\left(  {\mathbf{{r}},\omega}\right)  }}{{c\sqrt{\epsilon_{m}%
}|E^{\left(  0\right)  }|^{2}}}. \label{eq16}%
\end{equation}

\section{Optical parameters of an ensemble of spheroidal metal nanoparticles}

Below, an ensemble of spheroidal (ellipsoid of rotation) metal nanoparticles
is considered. This shape is the simplest among asymmetric ones, because its
asymmetry is characterized by a single dimensionless parameter: either the
ratio between the spheroid curvature radii , $\rho=R_{\bot}/R_{||}$, or the
spheroid eccentricity $e_{p}=\sqrt{\left\vert {1-\rho^{2}}\right\vert }$. In
this case, the obtained theoretical results can be applied to explain the
optical properties of particles within a wide interval of shapes, from discoid
to rod-like, by simply changing the curvature radii.

First of all, we are interested in how the dispersion of nanoparticle shapes
affects the optical characteristics of nanoparticle ensemble. Recall that the
nanoparticle form governs the frequencies and the number of plasma resonances.
Therefore, in order to emphasize the shape effect, let us assume that the
nanoparticles in the ensemble have the same volume $V$, but different
eccentricities $e_{p}$'s.

The coefficient of electric light absorption by a single spheroidal metal
nanoparticle can be obtained as the sum
\begin{equation}
K\left(  {\omega,\rho}\right)  =\,K_{\bot}\left(  {\omega,\rho}\right)
+K_{||}\left(  {\omega,\rho}\right)  , \label{eq17}%
\end{equation}
where each $s$-th component $(s=\bot,\Vert)$ is determined by Eq.~(\ref{eq16}%
), so that
\begin{multline}
K_{s=\bot,\Vert}\left(  {\omega,\rho}\right) \label{17a}\\
=\dfrac{\mu_{s}\,\omega^{4}\sigma_{s}\left(  {\omega,\rho}\right)  }{\left[
(\omega^{2}-\omega_{s}^{2})g_{s}(\rho)\right]  ^{2}+\left[  4\pi\,L_{s}%
(\rho)\sigma_{s}(\omega,\rho)\omega\right]  ^{2}}.
\end{multline}
Here, the averaging over the orientations of the spheroid symmetry axis has
already been made, and the following notations are introduced:%
\begin{equation}%
\begin{array}
[c]{l}%
\mu_{||}=\dfrac{4\pi\epsilon_{m}^{3/2}}{3c},\quad\mu_{\bot}=2\mu_{||},\\
g_{s}(\rho)=\epsilon_{m}+L_{s}(\rho)\,(1-\epsilon_{m});
\end{array}
\label{eq18}%
\end{equation}
the quantities%
\[
\omega_{s}(\rho)=\omega_{p}\sqrt{\frac{L_{s}(\rho)}{g_{s}(\rho)}}%
\]
are the frequencies corresponding to collective (plasma) oscillations of
conduction electrons in the nanoparticles in the directions perpendicularly
$(s=\perp)$ and in parallel $(s=\Vert)$ to the spheroid symmetry axis; and
$\sigma_{\bot,\parallel}(\omega,\rho)$ are the components of the optical
conductivity tensor.

In the coordinate frame with the axis $OZ$ directed along the spheroid
symmetry axis, the diagonal components of the optical conductivity tensor look
like
\begin{equation}
\sigma_{xx}=\sigma_{yy}\equiv\sigma_{\bot},\quad\sigma_{zz}\equiv
\sigma_{\parallel}. \label{eq20}%
\end{equation}
Analogously,
\begin{equation}
L_{x}=L_{y}\equiv L_{\bot},\quad L_{z}=L_{\parallel}.
\end{equation}
The principal values of the depolarization tensor are
\begin{align}
L_{\bot}  &  =\frac{1}{2}(1-L_{||}),\label{eq21a}\\
L_{||}  &  =\left\{  {%
\begin{array}
[c]{ll}%
\dfrac{1-e_{p}^{2}}{2e_{p}^{3}}\left[  \ln\left(  {\dfrac{1+e_{p}}{1-e_{p}}%
}\right)  -2e_{p}\right]  & \mathrm{for}~\rho<1,\\
\dfrac{1+e_{p}^{2}}{e_{p}^{3}}(e_{p}-\arctan e_{p}) & \mathrm{for}~\rho>1.
\end{array}
}\right.  \label{eq21b}%
\end{align}

In Refs~\cite{2,5}, it was shown that, if a non-spherical metal nanoparticle
is smaller than the electron free path length in it, the optical conductivity
becomes a tensor quantity, unlike the spherical case, where it is a scalar.
The components of this tensor were also analyzed in Refs~\cite{2,5} in the
general case of ellipsoidal nanoparticles and in various limiting cases. Here,
we are interested in the non-zero components of the optical conductivity
tensor for spheroidal nanoparticles, $\sigma_{\bot}$ and $\sigma_{||}$. Let us
confine the consideration to the case when the influence of the nanoparticle
shape on the optical conductivity tensor components is maximum. Expression
(\ref{eq10}) makes allowance for both the bulk (by means of the parameter
$\nu$) and the surface (by means of the characteristic $t_{0}$) electron
scattering. The mere surface electron scattering can be formally obtained from
Eq.~(\ref{eq10}) by putting $\nu\rightarrow0$. Actually, we mean the
inequality $\nu\ll v_{F}/R$, i.e. the bulk collision frequency is small in
comparison with the frequency of electron oscillations between the
nanoparticle walls. As a result, we obtain (for more details, see
Refs~\cite{2,5}) that%
\begin{widetext}%
\begin{equation}
\sigma_{\bot}=\frac{3\sigma_{0}}{16}\times\left\{  {%
\begin{array}
[c]{ll}%
\dfrac{\left(  {1-e_{p}^{2}}\right)  ^{1/3}}{e_{p}^{3}}\left[  {e_{p}\left(
{1+2e_{p}^{2}}\right)  -}\dfrac{{\left(  {1-4e_{p}^{2}}\right)  }}{{\left(
{1-e_{p}^{2}}\right)  ^{1/2}}}{\arcsin e_{p}}\right]   & \mathrm{for}%
~\rho<1,\\
\dfrac{\left(  {1+e_{p}^{2}}\right)  ^{1/3}}{e_{p}^{3}}\left[  {-e_{p}\left(
{1-2e_{p}^{2}}\right)  +}\dfrac{{\left(  {1+4e_{p}^{2}}\right)  }}{{\left(
{1+e_{p}^{2}}\right)  ^{1/2}}}{\ln\left(  {e_{p}+\sqrt{1+e_{p}^{2}}}\right)
}\right]   & \mathrm{for}~\rho>1,
\end{array}
}\right.  \label{eq22}%
\end{equation}%
\begin{equation}
\sigma_{||}=\frac{3\sigma_{0}}{8}\times\left\{  {%
\begin{array}
[c]{ll}%
\dfrac{\left(  {1-e_{p}^{2}}\right)  ^{1/3}}{e_{p}^{3}}\left[  {-e_{p}\left(
{1-2e_{p}^{2}}\right)  +}\dfrac{1}{{\left(  {1-e_{p}^{2}}\right)  ^{1/2}}%
}{\arcsin e_{p}}\right]   & \mathrm{for}~\rho<1,\\
\dfrac{(1+e_{p}^{2})^{^{1/3}}}{e_{p}^{3}}\left[  {e_{p}\left(  {1+2e_{p}^{2}%
}\right)  -}\dfrac{1}{{\left(  {1+e_{p}^{2}}\right)  ^{1/2}}}{\ln\left(
{e_{p}+\sqrt{1+e_{p}^{2}}}\right)  }\right]   & \mathrm{for}~\rho>1.
\end{array}
}\right.  \label{eq23}%
\end{equation}%
\end{widetext}%

\noindent Here,
\begin{equation}
\sigma_{0}=\frac{3n_{0}e^{2}}{2m\omega^{2}}\nu_{s}=\frac{n_{0}e^{2}}%
{m\omega^{2}}\left(  {\frac{3}{4}\frac{v_{F}}{R}}\right)  \label{eq25}%
\end{equation}
is the well-known expression for the optical conductivity of spherical
nanoparticle, in which
\begin{equation}
\nu_{s}=\frac{{v_{F}}}{2R}\label{eq24}%
\end{equation}
is the electron oscillation frequency between the walls, $v_{F}$ is the Fermi
velocity, and $R$ the radius of spherical particle. For an ellipsoidal
nanoparticle, the parameter $\nu_{s}$ is the electron oscillation frequency
between the walls but in a spherical nanoparticle with the same volume. One
can see that the both results (\ref{eq22}) and (\ref{eq23}) tends to formula
(\ref{eq25}), if the nanoparticle shape approaches the spherical one (at
$e_{p}\rightarrow0$). In expressions (\ref{eq22}) and (\ref{eq23}), we omitted
the oscillatory terms associated with the frequency resonance between the
external electromagnetic wave and the electron oscillations between the walls.
In the visible spectral interval, this resonance is not significant (for more
details, see Refs.~\cite{13,14}).

Therefore, the principal values of not only the depolarization tensor [see
Eqs.~(\ref{eq21a}) and (\ref{eq21b})], but also of the optical conductivity
tensor [Eqs.~(\ref{eq22}) and (\ref{eq23})], depend on the nanoparticle shape
described by the eccentricity $e_{p}$ (or the curvature radius ratio $\rho$).
Therefore, the both dependencies have to be taken into account when averaging
over the shape spread.

\begin{figure}[tb]
\centering  \includegraphics[width=\figw]{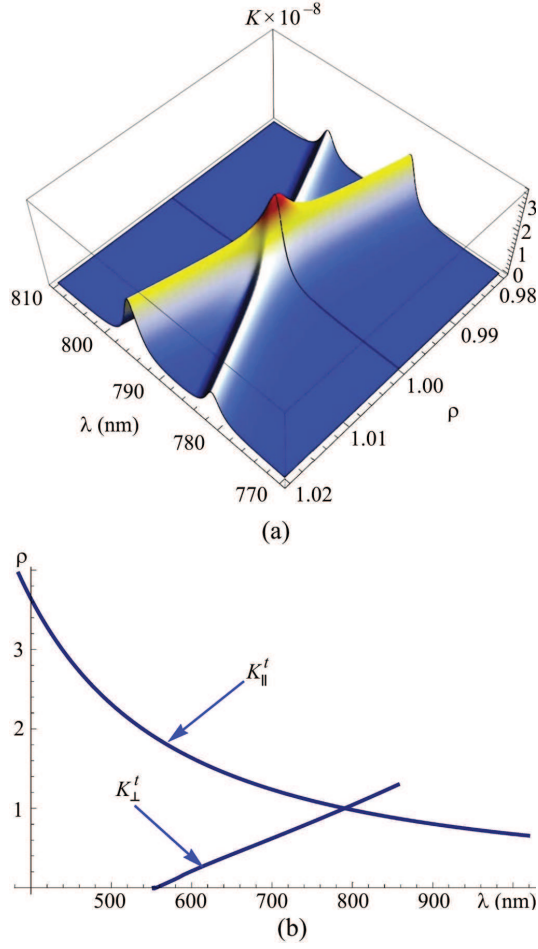}
\caption{(Color online) (a)~Dependence of the
coefficient $K$ of electric light absorption by a spheroidal nanoparticle
averaged over the nanoparticle orientation on the electromagnetic wave length
$\lambda$ and the nanoparticle shape described by the curvature radius ratio
$\rho$. (b)~Traces of plasma resonances $K_{\bot}^{t}$ and $K_{||}^{t}$ in the
plane $\left(  {\lambda,\rho}\right)  $.}%
\label{fig1}%
\end{figure}

In Fig.~\ref{fig1}(a), the dependence $K\left(  {\lambda,\rho}\right)  $ of
the coefficient $K$ of electric light absorption by a spheroidal nanoparticle
averaged over the nanoparticle orientation on the electromagnetic wave length
$\lambda$ and the nanoparticle shape described by the curvature radius ratio
$\rho$ is depicted. Two ridges correspond to two plasma resonances, which
mutually intersect at the $\left(  {\lambda,\rho}\right)  $-point with
$\rho=1$ corresponding to the plasma resonance in a spherical nanoparticle.
Panel~(a) only exhibits a local fragment of the dependence $K\left(
{\lambda,\rho}\right)  $. For a more complete understanding of the behavior of
indicated plasma resonances, panel~(b) illustrates their traces $K_{\bot}^{t}$
and $K_{||}^{t}$ in a wider area of the plane $\left(  {\lambda,\rho}\right)
$.

In order to obtian an expression for the total light absorption coefficient
for an ensemble of metal nanoparticles that takes into account the spread of
particle shapes, expression (\ref{eq17}) has to be averaged with the weigh
$\Pi(\rho)$. The latter describes the probability to find a nanoparticle with
the given $\rho$-value in the ensemble. We assume the function $\Pi(\rho)$ to
be normalized to the nanoparticle concentration $N$:%
\begin{equation}
\int\limits_{0}^{\infty}{\Pi\left(  \rho\right)  }\,d\rho=N, \label{eq26}%
\end{equation}
In terms of the function $P(\rho)=\frac{1}{N}{\Pi\left(  \rho\right)  }$, the
apparent value of the total absorption coefficient reads%
\begin{equation}
\overline{K\left(  \omega\right)  }=\int\limits_{0}^{\infty}{K\left(
{\omega,\rho}\right)  P\left(  \rho\right)  }\,d\rho. \label{eq27}%
\end{equation}

\section{Selection of the function $P(\rho)$}

The function $P(\rho)$ is nothing else but a probability to find a
nanoparticle with a definite $\rho$-value from the interval $0\leq\rho<\infty
$. It can be considered as the probability density for the function $K\left(
{\omega,\rho}\right)  $, so that Eq.~(\ref{eq27}) describes the averaging of
the total absorption coefficient of the nanoparticle ensemble over the
nanoparticle shapes.

While constructing $P(\rho)$, let us firstly pay attention that its domain
includes only non-negative $\rho$-values, so that
\begin{subequations}
\label{28}%
\begin{equation}
P(\rho<0)=0.
\end{equation}
Let the distribution $P(\rho)$ be characterized by a maximum at a certain
$\rho$-value $\rho=a$, which separates the regions of \textquotedblleft
oblate\textquotedblright\ ($0<\rho<a$) and \textquotedblleft
prolate\textquotedblright\ ($\rho>a$) particles. Let the shape distribution of
\textquotedblleft prolate\textquotedblright\ particles ($R_{\bot}\geq aR_{||}%
$) be described by the Gaussian function%
\begin{equation}
P(\rho>a)=\alpha\exp[-\beta_{1}(\rho-a)^{2}]\,, \label{28b}%
\end{equation}
where the parameters $\alpha$ and $\beta_{1}$ have a clear meaning.

The expression for the function $P(\rho)$ within the \textquotedblleft
oblate\textquotedblright\ interval $0\leq\rho\leq a$ was chosen to satisfy the
continuity conditions at its ends, i.e. $P(\rho=0)=0$ and $P(\rho
=a+0)=P(\rho=a-0)=\alpha$. For this purpose, we selected the function%
\begin{equation}
P(0\leq\rho\leq a)=\alpha\exp\left[  {-\frac{\beta_{0}\,(\rho
-a)^{2}}{a^{2}-(\rho-a)^{2}}}\right]  ,\label{28c}%
\end{equation}
which is an extention of Sobolev's \textquotedblleft hat\textquotedblright%
\ function \cite{15,16}. The meaning of the parameters $\alpha$ and $\beta
_{0}$ in Eq.~(\ref{28c}) is also quite clear. The properties of Sobolev's
\textquotedblleft hat\textquotedblright\ function make it very attractive for
the solution of a good many problems \cite{17}. An additional advantage of
this choice is a high smoothness of the resulting \textquotedblleft
cap\textquotedblright\ function (\ref{28}) within the whole interval of its
definition, because both Gaussian (\ref{28b}) and function (\ref{28c}) are
infinitely differentiable ones. The value of the parameter $\alpha$
corresponds to the maximum value of the function $P(\rho)$ at $\rho=a$ and is
determined from the normalization condition $\int\limits_{0}^{\infty}%
{P(\rho)\,d\rho}=1$.

The proposed model function $P(\rho)$ was selected on the heuristic principle.
Its application strongly simplifies the solution of a rather complicated
problem concerning the influence of the metal nanoparticle shape dispersion on
the total absorption coefficient of nanoparticle ensemble. The functions
describing the size distribution of nanoparticles \cite{18, 19} are an
argument in favor of our choice.

The illustrative calculations below were made for three parameter sets (I, II,
and III) of the function ${P(\rho)}$, which are quoted in Table~\ref{tab1}.
The corresponding plots are depicted in Fig.~\ref{fig2}.

\begin{table}[tb]
\caption{Parameter sets of the function $P(\rho)$.}%
\label{tab1}
\begin{ruledtabular}
\begin{tabular}{cccccc}
Variant & $a$ & $\beta_{0}$ & $\beta_{1}$ & $\alpha$\\ \hline
I & 1.0 & 1.0 & 1.638 & 0.772\\
II & 1.0 & 0.194 & 0.262 & 0.391\\
III & 2.0 & 1.0 & 0.410 & 0.386
\end{tabular}
\end{ruledtabular}
\end{table}

\begin{figure}[tb]
\centering  \includegraphics[width=\figw]{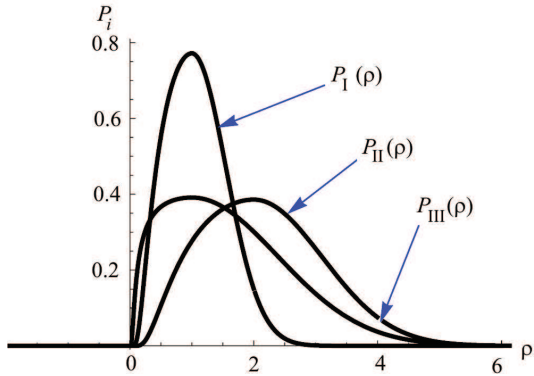}
\caption{(Color online) ``Cap'' probability
functions $P_{\mathrm{I,II,III}}(\rho)$ describing the distributions of
nanoparticles over their shape.}%
\label{fig2}
\end{figure}

\section{Results of computational experiment and their interpretation}

In computations, the following values of problem parameters{}were used:
$\omega_{p}=1.37\times10^{16}~\mathrm{s}^{-1}$, $\nu=3.39\times10^{13}%
~\mathrm{s}^{-1}$, $R=2.0\times10^{-6}~\mathrm{cm}$, $c=3.0\times
10^{10}~\mathrm{cm/s}$, $v_{F}=1.39\times10^{8}~\mathrm{cm/s}$, $n=10^{22}%
~\mathrm{cm}^{-3}$, $m=9.11\times10^{-27}~\mathrm{g}$, and $\epsilon_{m}=16.$

One of the results obtained for the dependence $K\left(  {\lambda,\rho
}\right)  $ of the absorption coefficient for a spheroidal metal nanoparticle
is shown in Fig.~\ref{fig1}(a). This figure illustrates the dependence
$K\left(  {\lambda,\rho}\right)  $ in the vicinity of resonance with the
external electromagnetic field for nanoparticles which forms are close to the
spherical one ($\rho=1$). Let us also introduce two functions, $K_{\bot
}^{(\max)}(\rho)$ and $K_{||}^{(\max)}(\rho)$, corresponding to the first and
second components, respectively, in Eq.~(\ref{eq17}). They are projections of
the maximum values of the orthogonal ($\perp$) and parallel ($\parallel$)
components of the light absorption coefficient $K\left(  {\omega,\rho}\right)
$ for the spheroidal metal nanoparticle onto the plane $(K,\rho)$ [see
Fig.~\ref{fig3}(a)]. (In what follows, it is more convenient to use the
frequency dependence of sought coefficients, taking into account the
dependences of the plasma resonance frequencies on the nanoparticle shape.) In
other words, $K_{\bot}^{(\max)}(\rho)$ and $K_{||}^{(\max)}(\rho)$ are
projections of two ridges in the dependence $K\left(  {\omega,\rho}\right)  $
corresponding to two plasma resonances (Fig.~\ref{fig1}) onto the plane
$(K,\rho)$ [Fig.~\ref{fig3}(a)]. Three segments in Fig.~\ref{fig3}(a)
illustrates the half-widths of the functions $P_{i}(\rho)$ ($i=\mathrm{I}%
,\mathrm{II},\mathrm{III}$). The $x$-coordinates $\rho_{a}$ and $\rho_{b}$ of
segment ends coincide with the calculated values $\rho_{a}^{(\mathrm{I}%
)}\approx0.36$, $\rho_{b}^{(\mathrm{I})}\approx1.65$, $\rho_{a}^{(\mathrm{II}%
)}\approx0.13$, $\rho_{b}^{(\mathrm{II})}\approx2.63$, $\rho_{a}%
^{(\mathrm{III})}\approx0.72$, and $\rho_{b}^{(\mathrm{III})}\approx3.30$. The
ordinate positions of those segments are irrelevant.

\begin{figure}[tb]
\centering \includegraphics[width=\figw]{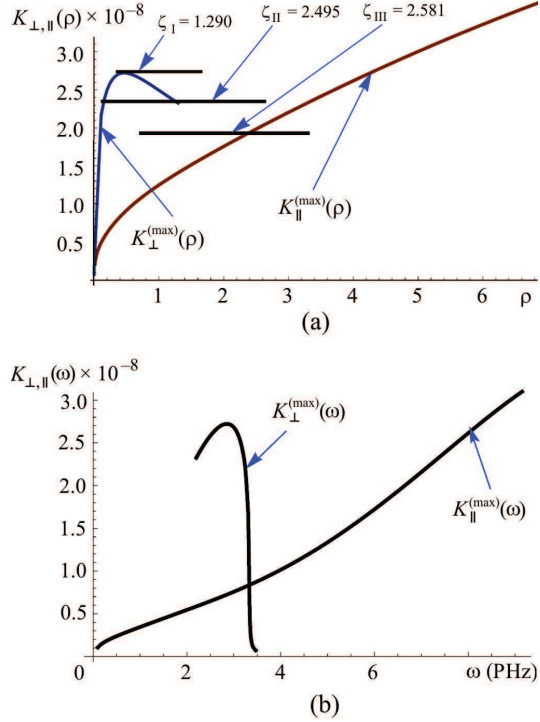}
\caption{(Color online) (a)~Projections $K_{\bot
}^{(\max)}(\rho)$ and $K_{||}^{(\max)}(\rho)$ of the maximum values of the
orthogonal and parallel components of the absorption coefficient $K\left(
{\omega,\rho}\right)  $ for the spheroidal metal nanoparticle onto the plane
$(K,\rho)$, and the segments corresponding to the half-widths of the functions
$P_{i}(\rho)$ $(i=\mathrm{I\,,II\,,III})$. (b)~The same as in panel~(a), but
the projections $K_{\bot}^{(\max)}(\omega)$ and $K_{||}^{(\max)}(\omega)$ are
made onto the plane $(K,\omega)$.}%
\label{fig3}
\end{figure}

Similarly, the functions $K_{\bot}^{(\max)}(\omega)$ and $K_{||}^{(\max
)}(\omega)$ are the projections of the maximums of the orthogonal and
parallel, respectively, components of the absorption coefficient by a
spheroidal metal nanoparticle onto the plane $(K,\omega)$ [Fig.~\ref{fig3}(b)].

First of all, attention is attracted by the fact that the domain of the
orthogonal component (ridge) of the light absorption coefficient, $K_{\bot
}\left(  {\omega,\rho}\right)  $, is very narrow. Really, as follows from
calculations, the function $K_{\bot}^{(\max)}(\rho)$ is determined in the
interval $\rho\in(0,1.414)$ [Fig.~\ref{fig3}(a)], whereas the domain of the
function $K_{\bot}^{(\max)}(\omega)$ is the interval $\omega\in(2.14\times
10^{15}~\mathrm{s}^{-1},3.47\times10^{15}~\mathrm{s}^{-1})$ [Fig.~\ref{fig3}%
(b)]. On the other hand, from the dependence of the plasma resonance frequency
on the nanoparticle shape, it follows that the transverse resonance frequency
$\omega_{\bot}(\rho)$ falls within the interval $(2.14\times10^{15}%
~\mathrm{s}^{-1},3.32\times10^{15}~\mathrm{s}^{-1})$. (We conventionally call
plasma oscillations in the directions along the spheroid axis and
perpendicularly to it as longitudinal and transverse, respectively.) Thus, the
domain of the function $K_{\bot}^{(\max)}(\omega)$ is composed of two
sections: the resonance, $\omega_{r}\in(2.14\times10^{15}~\mathrm{s}%
^{-1},3.32\times10^{15}~\mathrm{s}^{-1})$, and the deflation, $\omega_{d}%
\in\lbrack3.32\times10^{15}~\mathrm{s}^{-1},3.47\times10^{15}~\mathrm{s}%
^{-1})$, ones. In the latter, $K_{\bot}^{(\max)}(\omega)$ drastically
decreases to the background values of $K\left(  {\omega,\rho}\right)  $ (In
geology, \textit{deflation} is the process of the particle dispersing and
removal by the wind.) It is also important to mark that there is a single
plasma resonance in $K_{||}\left(  {\omega,\rho}\right)  $ at $\omega
>3.47\times10^{15}~\mathrm{s}^{-1}$.

The main aim of the computational experiment was to determine the apparent
values of the total absorption coefficient,%
\end{subequations}
\begin{equation}
\overline{K_{i}\left(  \omega\right)  }=\int\limits_{0}^{\infty}{K\left(
{\omega,\rho}\right)  P_{i}\left(  \rho\right)  }\,d\rho\;, \label{eq38}%
\end{equation}
for various functions $P_{i}(\rho)$ $(i=\mathrm{I,\,II,\,III})$. The
corresponding results are shown in Fig.~\ref{fig4}.

\begin{figure}[ptb]
\centering \includegraphics[width=\figw]{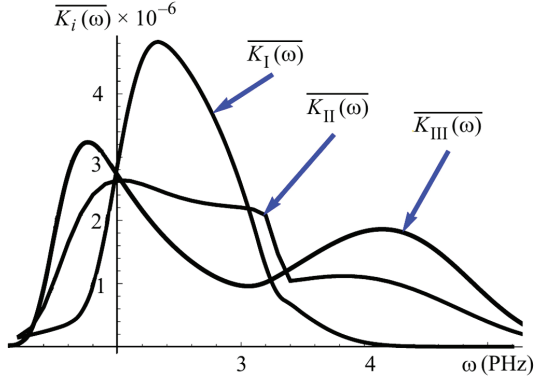}
\caption{(Color online) Averaged dependences{}%
of the total absorption coefficient $\overline{K\left(  \omega\right)  }$ for
ensembles of spheroidal nanoparticle with various nanoparticle-shape
distribution functions ${P_{i}\left(  \rho\right)  }$ ($i=\mathrm{I,II,III}%
$).}%
\label{fig4}%
\end{figure}

By comparing the spectral dependence of the total light absorption coefficient
for a single spheroidal metal nanoparticle (Fig.~\ref{fig1}) with the same
dependence for nanoparticle ensembles (Fig.~\ref{fig4}), we arrive at a
conclusion that the corresponding values differ from one another by about two
orders of magnitude. This ratio is rather general. Let us consider the plots
of the functions $\overline{K_{i}\left(  \omega\right)  }$ (Fig.~\ref{fig4})
in more detail.

A comparison between the plots of the functions $K_{\bot}^{(\max)}(\rho)$ and
$K_{||}^{(\max)}(\rho)$, on the one hand, and the half-widths $\zeta_{i}$ of
the functions $P_{i}(\rho)$ $(i=\mathrm{I\,,II\,,III})$, on the other hand,
brings us to the following conclusions.

$\bullet$~The main factor \textquotedblleft affecting\textquotedblright\ the
function $P_{I}(\rho)$ at integral transformation (\ref{eq38}) is the
orthogonal component $K_{\perp}\left(  {\omega,\rho}\right)  $ of the
coefficient of light absorption by a spheroidal metal nanoparticle $K\left(
{\omega,\rho}\right)  $ [cf. $K_{\bot}^{(\max)}(\rho)$ and $\zeta_{I}$ in
Fig.~\ref{fig3}(a)]. Therefore, in the resonance region $\omega_{r}$, the
curve $\overline{K_{I}\left(  \omega\right)  }$ in Fig.~\ref{fig4} is, in a
sense, similar to its \textquotedblleft parent\textquotedblright\ $P_{I}%
(\rho<1.414)$. The behavior of the curve $\overline{K_{I}\left(
\omega\right)  }$ in Fig.~\ref{fig4} appreciably changes in the deflation
section $\omega_{d}$.

$\bullet$~The function $\overline{K_{II}\left(  \omega\right)  }$ changes the
most substantially in the deflation section $\omega_{d}$ and its right wing,
where the influence of $K_{||}^{(\max)}(\rho>1.414)$ and $K_{||}^{\max}%
(\omega>3.47\times10^{15}~\mathrm{s}^{-1})$ is appreciable.

$\bullet$~The function $\overline{K_{III}\left(  \omega\right)  }$ most
adequately describes the real situation. It completely takes into account the
influence of both the transversal $K_{\bot}\left(  {\omega,\rho}\right)  $ and
longitudinal $K_{||}\left(  {\omega,\rho}\right)  $ components of the light
absorption coefficient on $P_{III}(\rho)$.

\section{Magnetic absorption}

Let us consider the frequency interval%
\begin{equation}
\nu\ll\omega\ll\omega_{s}\quad(s=\bot,||). \label{eq39}%
\end{equation}
For typical metals, $\nu\sim10^{13}~\mathrm{s}^{-1}$. The inequality $\nu
\ll\omega$ allows us to neglect the bulk scattering of electrons. The
inequality $\omega\ll\left(  \omega_{\bot},\omega_{||}\right)  $ means that we
are far from plasma resonances, and the electrical absorption does not
\textquotedblleft obscure\textquotedblright\ the magnetic one . By
substituting Eq.~(\ref{eq13}) into Eq.~(\ref{eq15}), we obtain the following
expression for the energy of magnetic light absorption by a spheroidal
($R_{1}=R_{2}\equiv R_{\bot}$, $R_{3}\equiv R_{||}$) metal nanoparticle
\cite{5, 22}:%
\begin{multline}
W_{m}=\frac{9}{128}V\frac{n\,\,e^{2}}{mc^{2}}v_{F}R_{\bot}\label{eq40}\\
\times\left[  {\chi_{H}\left(  {H_{||}^{(0)}}\right)  ^{2}+\eta_{H}\left(
{\frac{R_{||}^{2}}{R_{\bot}^{2}+R_{||}^{2}}H_{\bot}^{(0)}}\right)  ^{2}%
}\right]  .
\end{multline}
Here, $H_{||}^{(0)}$ and $H_{\bot|}^{(0)}$ are the amplitudes of the magnetic
wave field components that are parallel ($\parallel$) and orthogonal ($\perp$)
to the rotation axis of spheroid. Formula (\ref{eq40}) includes only the
magnetic components, because they are engaged in the magnetic light absorption
by a nonspherical nanoparticle. Finally,%

\begin{widetext}%
\begin{equation}
\chi_{H}=\dfrac{1}{8e_{p}^{2}}\times\left\{  {%
\begin{array}
[c]{ll}%
\left(  {1+2e_{p}^{2}}\right)  \left(  {1-e_{p}^{2}}\right)  ^{1/2}%
-\dfrac{{1-4e_{p}^{2}}}{e_{p}}\arcsin e_{p} & \mathrm{for~}\rho<1,\\
-\left(  {1-2e_{p}^{2}}\right)  \left(  {1+e_{p}^{2}}\right)  ^{1/2}%
+\dfrac{{1+4e_{p}^{2}}}{e_{p}}\ln\left(  {e_{p}+\sqrt{1+e_{p}^{2}}}\right)  &
\mathrm{for~}\rho>1;
\end{array}
}\right.  \label{eq41}%
\end{equation}%
\begin{equation}
\eta_{H}=\frac{1}{4e_{p}^{2}}\times\left\{  {%
\begin{array}
[c]{ll}%
-\left(  {1-8e_{p}^{2}+4e_{p}^{4}}\right)  \left(  {1-e_{p}^{2}}\right)
^{1/2}+\dfrac{{1+2e_{p}^{2}}}{e_{p}}\arcsin e_{p} & \mathrm{for~}\rho<1,\\
\left(  {1+8e_{p}^{2}+4e_{p}^{4}}\right)  \left(  {1+e_{p}^{2}}\right)
^{1/2}-\dfrac{{1-2e_{p}^{2}}}{e_{p}}\ln\left(  {e_{p}+\sqrt{1+e_{p}^{2}}%
}\right)  & \mathrm{for~}\rho>1.
\end{array}
}\right.  \label{eq42}%
\end{equation}%
\end{widetext}%

\noindent In the case of spherical nanoparticle, $R_{\bot}=R_{||}=a$, i.e.
$e_{p}\rightarrow0$, so that $\chi_{H}=2/3$ and $\eta_{H}=8/3$, and we obtain
the result of Ref.~\cite{5},%
\begin{equation}
W_{m}^{(0)}=\frac{3}{64}V\frac{n\,\,e^{2}v_{F}}{mc^{2}}a\left(  {H^{(0)}%
}\right)  ^{2}. \label{eq43}%
\end{equation}

In what follows, when studying the dependence of absorption by a nanoparticle
on its shape, it is convenient to consider the ratio between the energy
absorbed by a spheroidal particle and the energy absorbed by a spherical
particle with the same volume, i.e. the ratio of expressions (\ref{eq40}) and
(\ref{eq43}):%
\begin{equation}
\frac{W_{m}}{W_{m}^{(0)}}=\frac{3}{2}\rho^{1/3}\frac{\chi_{H}\left(
{H_{||}^{(0)}}\right)  ^{2}+\dfrac{\eta_{H}}{(1+\rho^{2})^{2}}\left(
{H_{\bot}^{(0)}}\right)  ^{2}}{\left(  {H^{(0)}}\right)  ^{2}}. \label{eq44}%
\end{equation}
When deriving this formula, we took into account that%
\[
\frac{R_{\bot}}{a}=\frac{R_{\bot}}{(R_{\bot}^{2}R_{||})^{1/3}}=\rho^{1/3}.
\]
Ratio (\ref{eq44}) is also equal to the ratio between the absorption
coefficients of spheroidal and spherical nanoparticles. In the coordinate
frame oriented along the principal spheroid axes,%
\begin{equation}
H_{||}^{(0)}=H^{(0)}\cos\theta,\quad H_{\bot}^{(0)}=H^{(0)}\sin\theta,
\label{eq45}%
\end{equation}
where $\theta$ is the angle between the vector $\mathbf{H}^{(0)}$ and the
spheroid axis of rotation. As a result, we obtain%
\begin{equation}
\frac{W_{m}}{W_{m}^{(0)}}=\frac{3}{2}\rho^{1/3}\left[  \chi_{H}\cos^{2}%
\theta+\frac{\eta_{H}}{(1+\rho^{2})^{2}}\sin^{2}\theta\right]  . \label{eq46}%
\end{equation}

\begin{figure}[tb]
\centering \includegraphics[width=\figw]{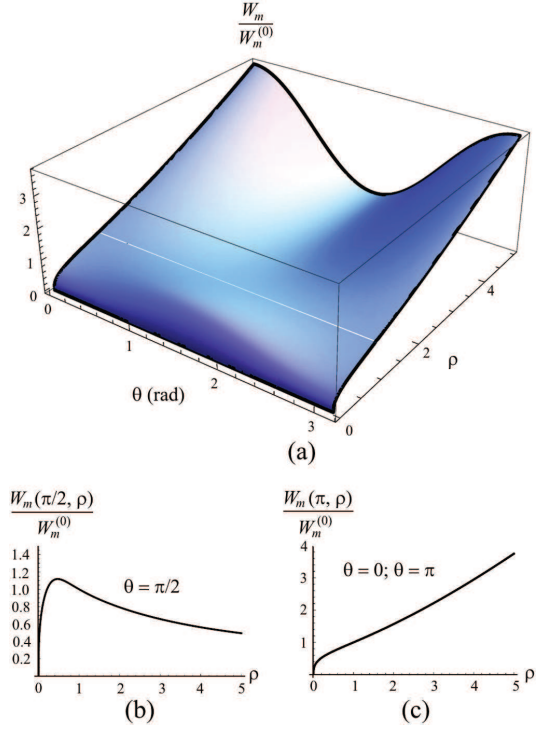}
\caption{(Color online) (a)~3D
plot demonstrating the dependence of the magnetic energy $W_{m}(\rho,\theta)$
absorbed by a spheroidal particle on its shape (the parameter $=R_{\bot
}/R_{||}$) and orientation with respect to the electromagnetic wave vector
(the angle $\theta$). The values are normalized by the energy $W_{m}^{(0)}$
absorbed by a spherical particle. (b and c)~Profiles of this surface along
(b)$~\theta=\pi/2$ and (c)~$\theta=0$ and $\pi$.}%
\label{fig5}%
\end{figure}

A 3D plot demonstrating the dependence of this ratio on the variables $\rho$
and $\theta$ is shown in Fig.~\ref{fig5}. The adequacy of the plotted
geometric surface to the physical picture is evidenced by the following facts.
First of all, attention is drawn by a boundary between two surface sections,
convex (at $0\leq\rho<1$) and concave (at $\rho>1$) ones. This boundary is a
straight line, with its every point satisfying the equality%
\[
W_{m}(\rho=1,0\leq\theta\leq\pi)=W_{m}^{(0)}.
\]
From the physical viewpoint, this equality is confirmed by a simple fact: for
a particle with $\rho=R_{\bot}/R_{||}=1$, the absorbed energy is identical to
that adsorbed by a spherical particle, $W_{m}(\rho=1,\theta)=W_{m}^{(0)}$,
irrespective of the angle $\theta$ between the vector $\mathbf{H}^{(0)}$ and
the axis of sphere rotation.

Another interesting fact is the growth of the energy absorbed by a spheroidal
nanoparticle as the ratio $\rho=R_{\bot}/R_{||}$ between the curvature radii
increases (the growth of the disk-like character of nanoparticle shape) at any
angle $\theta$. The only $\theta$-value at which $W_{m}(\rho,\theta)$
asymptotically approaches zero at $\rho\rightarrow\infty$ is $\theta=\pi/2$
[Fig.~\ref{fig5}(b)]. At the same time, magnetic absorption by a spheroidal
metal nanoparticle symmetrically attains maximum at two $\theta$-values: 0 and
$\pi$ [Fig.~\ref{fig5}(b)]. Really, if an almost flat particle is oriented
orthogonally to the magnetic field vector ($\theta=0$ or $\pi$), then its
interaction with this field is maximum. But if the same (almost flat) particle
is oriented along the magnetic field vector ($\theta=\pi/2$), its interaction
with this field is small.

Now, let an ensemble of spheroidal metal nanoparticles be chaotically oriented
in a dielectric matrix. Then, expression (\ref{eq46}) should be averaged over
all values of the angle $\theta$. After averaging, the ratio between the
energies of magnetic light absorption by ensembles of randomly oriented
spheroidal and spherical (with the same volume) metal nanoparticle reads%
\begin{equation}
\left\langle \frac{W_{m}}{W_{m}^{(0)}}\right\rangle =\frac{3}{4}\rho
^{1/3}\left[  \chi_{H}+\frac{\eta_{H}}{(1+\rho^{2})^{2}}\right]  .
\label{eq47}%
\end{equation}

Note that magnetic absorption (\ref{eq47}) does not depend on the wave
frequency in interval (\ref{eq39}). This effect can be easily understood on an
example of spherical nanoparticles. Magnetic absorption is described by the
second term in expression (\ref{eq1}). In the frequency interval (\ref{eq39}),
as one can see from Eq.~(\ref{eq2}), we have $\omega{\epsilon}^{\prime\prime
}\sim\nu\omega_{p}^{2}/\omega^{2}$. If this relationship is substituted into
formula (\ref{eq1}), the frequency dependence of magnetic absorption
disappears. We should also emphasize that formula (\ref{eq1}) was obtained for
the case when the nanoparticle size strongly exceeds the electron free path
length. In the general case, i.e. at an arbitrary ratio between those
parameters, the problem of light scattering light by a spherical metal
particle was solved in Ref.~\cite{23}.

\begin{figure}[ptb]
\centering \includegraphics[width=\figw]{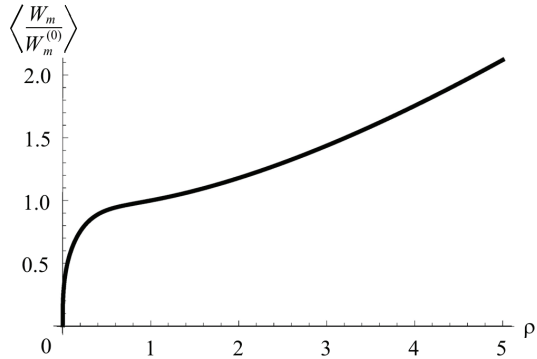}
\caption{(Color online) Dependence of the
average value of the ratio $\,W_{m}/W_{m}^{(0)}$ between the energies of
magnetic absorption by a spheroidal metal nanoparticle and a spherical
particle with the same volume on the curvature radius ratio $\rho=R_{\bot
}/R_{||}$ for the spheroidal particle.}%
\label{fig6}%
\end{figure}

Figure \ref{fig6} demonstrates the dependence of the ratio%
\[
\frac{\left\langle K_{m}\right\rangle }{K_{m}^{(0)}}=\left\langle \frac{W_{m}%
}{W_{m}^{(0)}}\right\rangle
\]
between the coefficients of magnetic light absorption by ensembles of
randomlyfig1 oriented spheroidal and spherical (with the same volume) metal
nanoparticle on the ratio $\rho=R_{\bot}/R_{||}$ between the spheroid
curvature radii (in the former ensemble). Here $K_{m}^{(0)}$ is the
coefficient of magnetic absorption by a spherical particle with the volume
equal to that of spheroidal particle. From this figure, the following
conclusion can be drawn: for spheroidal nanoparticles with more discoidal
shapes, the averaged magnetic absorption is higher. This growth has a smooth
character at $\rho>0.5$.

\section{Conclusions}

To summarize, a theory describing the dependence of electromagnetic energy
absorption by metal nanoparticles on their shape has been developed. Unlike
magnetic absorption, electric absorption was found to strongly depend on the
nanoparticle shape. This difference is explained by the fact that, in the
visible spectral interval, electric absorption is mainly associated with the
availability of plasma resonances in nanoparticles. However, the frequencies
of plasma resonances, their half-widths and number depend on the nanoparticle
shape. Therefore, by changing the parameter $\rho=R_{\bot}/R_{||}$ at a fixed
frequency $\omega$ of external irradiation, we can enter into the resonance
conditions with the characteristic plasma frequencies of nanoparticles and
exit from them. This circumstance is responsible for a drastic dependence of
electric absorption on the shape of metal nanoparticles. On the other hand, in
the frequency interval where it can be significant [see Eq.~(\ref{eq39})],
magnetic absorption does not depend on the frequency at all. As one can see
from Fig.~\ref{fig5}, the dependence of magnetic absorption by nanoparticles
on their shape mainly reveals itself in the angular dependence.

We presented the results of our theoretical studies concerning the optical
characteristics of an ensemble of spheroidal metal nanoparticle, such as the
components of the optical conductivity tensor, the principal values of
depolarization tensor components, and the absorption coefficient. The averaged
parameters were calculated taking into account the influence of the
nanoparticle shape on the depolarization coefficients and the optical
conductivity tensor components. The influence of nanoparticle shape on the
conductivity was taken into account in the averaging procedure for the first time.

It is well-known that the number of plasma resonances, their frequencies and
decrements depend on the nanoparticle shape. In particular, spherical
nanoparticles are characterized by one plasma resonance, spheroidal
nanoparticles by two, and elliptical ones by three plasma resonances. The
result of our calculation testifies that in the case of spheroidal
nanoparticle, there are two plasma resonances in a finite frequency interval
determined by the input problem parameters. Beyond this interval, only one
plasma resonance, $K_{||}\left(  {\omega,\rho}\right)  $, takes place.

A \textquotedblleft cap\textquotedblright\ function (a combination of the
\textquotedblleft hat\textquotedblright\ and Gaussian functions) was used to
approximate the distribution of nanoparticles over their shape, $P(\rho)$.
This model substantially simplifies the solution of a rather complicated
problem concerning the influence of the nanoparticle shape non-uniformity over
the ensemble on the total absorption coefficient. Furthermore, it was found to
be optimal for the qualitative theoretical study of the averaging over the
shape spread. The variation of the characteristic parameters of the function
$P(\rho)$, such as its half-width and the maximum position, was found to
substantially affect the averaged total absorption coefficient $\overline
{K\left(  \omega\right)  }$.


\begin{thebibliography}{99}                                                                                               %


\bibitem {1}V.V.~Klimov, \textit{Nanoplasmonics} (Fizmatlit, Moscow, 2010) (in Russian).

\bibitem {2}P.M. Tomchuk and N.I. Grigorchuk, Phys. Rev. B \textbf{73}, 155423 (2006).

\bibitem {3}E.F. Venger, A.V. Goncharenko, and M.L.~Dmytruk, \textit{Optics of
Small Particles and Dispersion Media} (Naukova Dumka, Kyiv, 1999) (in Ukrainian).

\bibitem {4}A.V. Goncharenko, E.F.~Venger, and S.N. Zavadskii, J. Opt. Soc.
Am. B \textbf{13}, 2392 (1996).

\bibitem {5}P.M. Tomchuk and B.P. Tomchuk, Zh. \`{E}ksp. Teor. Fiz.
\textbf{112}, 661 (1997).

\bibitem {6}R.D. Fedorovich, A.G. Naumovets, and P.M.~Tomchuk, Phys. Rep.
\textbf{328}, 73 (2000).

\bibitem {7}D.V. Butenko and P.M. Tomchuk, Surf. Sci. \textbf{606}, 1892 (2012).

\bibitem {8}C.F. Boren and D.R. Huffman, \textit{Absorption and Scattering of
Light by Small Particles} (John Wiley and Sons, New York, 1983).

\bibitem {9}H.C. van de Hulst, \textit{Light Scattering by Small Particles}
(John Wiley and Sons, New York, 1957).

\bibitem {10}L.D. Landau and E.M. Lifshits, \textit{Electrodynamics of
Continuous Media} (Pergamon Press, New York, 1984).

\bibitem {11}G. Mie, Beitr\"{a}ge zur Optik tr\"{u}ber Medien, speziell
kolloidaler Metall\"{o}sungen, \textit{Ann. Phys.} \textbf{25}, 377 (1908).

\bibitem {12}P.M. Tomchuk and D.V. Butenko, Int. J. Mod. Phys. B \textbf{31},
1750029 (2017).

\bibitem {13}N.I. Grygorchuk and P.M. Tomchuk, Ukr. Fiz. Zh. \textbf{51}, 921 (2006).

\bibitem {14}P.M. Tomchuk and D.V. Butenko, Ukr. Fiz. Zh. \textbf{60}, 1042 (2015).

\bibitem {15}S.L. Sobolev, \textit{Some Applications of Functional Analysis in
Mathematical Physics} (Nauka, Moscow, 1988) (in Russian).

\bibitem {16}V.S. Vladimirov, \textit{Equations of Mathematical Physics}
(Nauka, Moscow, 1988) (in Russian).

\bibitem {17}V.N. Starkov, M.S. Brodyn, P.M. Tomchuk V.Ya.~Gaivoronskyi, and
O.Yu. Boyarchuk, Ukr. Fiz. Zh. \textbf{60}, 602 (2015).

\bibitem {18}W. Haiss, N.T.K. Thanh, J. Aveyard, and D.G. Fernig, Anal. Chem.
\textbf{79}, 4215 (2007).

\bibitem {19}A. Carrillo-Cazares, N.P. Jim\'{e}nez-Mancilla, M.A.
Luna-Guti\'{e}rrez, K. Isaac-Oliv\'{e}, and M.A. Camacho-L\'{o}pez, J.
Nanomater. \textbf{2017}, 3628970 (2017).

\bibitem {22}M.I. Grygorchuk and P.M. Tomchuk, J. Phys. Stud. \textbf{9}, 135 (2005).

\bibitem {23}I.A. Kuznetsova, M.E. Lebedev, and A.A.~Yushkanov, Zh. Tekhn.
Fiz.\textit{ }\textbf{85}, N~9, 1 (2015).
\end{thebibliography}
\end{document}